\documentclass[twocolumn,aps,pra,showpacs]{revtex4}
\usepackage{amssymb}
\usepackage{amsmath}
\usepackage{graphicx}
\usepackage{epsfig}
\newcommand{\be}{\begin{equation}}
\newcommand{\ee}{\end{equation}}

\begin{document}

\title{Post-selected weak measurement beyond the weak value}

\author{Tam\'as Geszti}
\email{geszti@elte.hu}
\affiliation{Department of Physics of Complex Systems, E\"otv\"os University, 
Budapest, Hungary}
\begin{abstract}
Closed expressions are derived for the quantum measurement statistics 
of pre-and postselected gaussian particle beams. The weakness of the 
pre-selection step is shown to compete with the non-orthogonality of 
post-selection in a transparent way. The approach is shown to be useful 
in analyzing post-selection-based signal amplification, allowing 
measurements to be extended far beyond the range of validity of the 
well-known Aharonov-Albert-Vaidman limit. 
\end{abstract} 
\maketitle

Pre-and postselected weak measurements \cite{aav}, with their surprising
mathematical properties, often traced back to a weird combination of
Fourier components \cite{berry}, have raised considerable interest, strongly
supported by experimentally accessible predictions \cite{ritchie,pryde}. 
The standard discussion is centered about the so-called weak value
\be{}\label{1}
\langle\hat A\rangle_{weak}=\frac{\langle f|\hat A |i\rangle}
                   {\langle f | i \rangle}
\ee 
of an observable $\hat A$, $|i\rangle$ and $|f\rangle$ being two vectors 
- an initial and a final (``post-selected'') state - in the Hilbert space
on which $\hat A$ - say, a spin component of an atom - acts. Remaining with 
that example, the  Aharonov-Albert-Vaidman (AAV) procedure is a sophisticated 
two-step Stern-Gerlach-type measurement of that spin component. The first 
Stern-Gerlach separation, coupling the spin by magnetic interaction to the 
transverse position $z$ of the atomic beam, is weak, i.e. the resulting 
shifts $\pm d$ of the two beams are much less than the transverse width 
$w$ of each beam. That first step is followed by an appropriately oriented 
second - strong - Stern-Gerlach separation, marking out one of the branches
for further processing - that is called the post-selection. Finally 
\cite{von}, the measured value of $\hat A$ is read off through the 
statistics of atoms detected within the postselected branch at different
displacements $z$ \cite{detectors}. The full sequence can result in 
detecting displacements on the average much larger than $d$, being 
proportional to the real part of the weak value, Eq. (\ref{1}). A feasible 
alternative implementation of the scheme is to use light beams, polarization 
taking the place of spin, and birefringence - linear or circular - being used 
for the Stern-Gerlach-type separation. Most recently, several experiments 
\cite{kwiat,dixon} used the scheme to measure weak birefringence itself, 
giving rise to sub-nanometer-scale displacements; the AAV post-selection 
served for signal amplification of that \cite{ampl}. There was one change 
with respect to the scheme outlined above: the new measurements exploit 
the possibility of measuring the imaginary part of the weak value, which 
transforms the displacements into momentum space, with strongly enhanced 
amplification.

The aim of the present note is to analyze the situation in a straightforward
manner, as a two-beam interference phenomenon in which visible contrast is
created by the AAV sequence of weak pre-selection and strong post-selection.
The procedure can be analyzed in full generality; for simplicity, we present 
the relevant formulas for gaussian beams. In that context the AAV 
weak value, Eq. (\ref{1}), appears as a complex expression furnishing
measurable mean values in a well-defined limiting case. With that knowledge, 
practical tools are obtained to explore and analyze a wide range of 
experimentally accessible data far beyond that limiting case.

In the analysis that follows we find that although for $~d/w\to 0~$  
the measured mean value approaches the AAV weak value Eq. (\ref{1}), a small 
but {\sl finite} weakness ratio $~d/w~$ is competing in smallness with the 
distance from full ortogonality of the pre- and postselected states. We derive 
a closed expression for the measured mean value, valid for any degree of 
pre-selection weakness and post-selection non-orthogonality. Under the single 
condition of weakness, $~d/w\ll 1$, the formula is reduced to a simple 
non-linear function of $~d/w$, which if used for data fitting, can result 
in strongly enhanced signal amplification.

For the sake of concreteness, we speak of spin-1/2 particles prepared in
a definite initial spin state $\begin{pmatrix}\alpha\\\beta \end{pmatrix}$,
and the observable one wants to measure is the spin component $\hat\sigma_z$. 
The corresponding Stern-Gerlach separation, followed by free flight until 
reaching the detectors, produces the respective displacements $\pm d$ 
in the $z$ direction for the two spin components. We assume that the 
particles are propagating along the $y$ direction in gaussian beams, the 
width of which reaches the value $w$ in the plane of the detectors. 
Weak separation/measurement is defined by the condition $d\ll w$, 
i.e. the two beams remain strongly overlapping. Omitting trivial events 
in the other space dimensions, this first step of weak Stern-Gerlach 
separation can be described as
\be{}\label{2}
|i\rangle~=~\Phi(z)\begin{pmatrix}\alpha\\\beta \end{pmatrix}
\\\longrightarrow
       \begin{pmatrix}\alpha~\Phi(z-d)\\\beta~\Phi(z+d)\end{pmatrix},
\ee
where $\Phi(z)=\exp(-z^2/4w^2)/\sqrt{w\sqrt{2\pi}}$ is the
normalized transverse wave function of the beam in the direction $z$
of separation. Spin-orbital entanglement associates the two beams to 
orthogonal spin states, therefore straightforward detection would just 
add the intensities, to produce the probability distribution function
\be{}\label{3}
p(z)=|\alpha|^2~\Phi^2(z-d)+|\beta|^2~\Phi^2(z+d)
\ee
and the corresponding mean displacement
\be\label{4}
\langle z\rangle~=~(|\alpha|^2-|\beta|^2)~d,
\ee
which is practically invisible above the gaussian background of width $w$. 

However, the spin degree of freedom offers the possibility to create phase 
contrast by rotating the spin basis, which is subsequently exploited in the 
post-selection procedure (see below). The experimentally more accessible case 
of photons with two orthogonal states of polarization, and their spatial 
separation by one of the many effects giving rise to birefringence, then 
post-selecting by polarization rotation, makes no difference in the 
description, only the physics behind the displacement $d$ and the method
of preparation of pre- and postselected states change.

In the example at hand the post-selection step, to be carried out before
detection, consists in a second, this time strong Stern-Gerlach 
separation in a different direction $\xi$, rotated with respect to the 
initial one, fully separating the above state vector into two 
non-overlapping beams of the corresponding orthogonal spin states 
$\begin{pmatrix}\gamma\\\delta \end{pmatrix}$ and
$\begin{pmatrix}\delta^*\\ -\gamma^* \end{pmatrix}$:
 
\be{}\label{5}
\begin{split}
\longrightarrow~
&f_1(\xi)~\left(\alpha\gamma^*~\Phi(z-d)+\beta\delta^*~\Phi(z+d)\right) 
           \begin{pmatrix}\gamma\\\delta \end{pmatrix} 
\\+&f_2(\xi)~\left(\alpha\delta~\Phi(z-d)-\beta\gamma~\Phi(z+d)\right)
\begin{pmatrix}\delta^*\\ -\gamma^* \end{pmatrix}, 
\end{split}\ee
where $f_1(\xi)$ and $f_2(\xi)$ are two non-overlapping functions of the
post-selection coordinate $\xi$.

It is this step which creates visible phase contrast between the two 
$z-$shifted peaks, in each of the post-selected components. Any of them 
can be projected out by placing the detector into the corresponding band; 
e.g. for the first component: where $f_1(\xi)\not=0;~f_2(\xi)=0$. The 
final result is obtained by scanning with the detector along $z$, while
keeping $\xi$ unchanged. That is the final, non-unitary quantum measurement, 
the statistics of which is expected to furnish as output, through Born's 
rule, the fraction of incoming particles detected in the postselected 
channel, at displacement $z$:
\be{}\label{6}
\begin{split}
&q(z):=~\left|\alpha\gamma^*~\Phi(z-d)+\beta\delta^*~\Phi(z+d)\right |^2
\\&~=|\alpha|^2|\gamma|^2\Phi^2(z-d)+|\beta|^2|\delta|^2\Phi^2(z+d)
\\&~~~~~+2\Re (\alpha\beta^*\gamma^*\delta)e^{-\frac{d^2}{2w^2}}\Phi^2(z)
\end{split}\ee
where the last term is the contribution of visible interference, created
in the process of post-selection.

The analogous expression for the complementary channel, 
$f_2(\xi)\not=0;~f_1(\xi)=0$, reads  
$|\alpha|^2|\delta|^2\Phi^2(z-d)+|\beta|^2|\gamma|^2\Phi^2(z+d)
-2\Re (\alpha\beta^*\gamma^*\delta)e^{-\frac{d^2}{2w^2}}\Phi^2(z)$.
Adding that to Eq. (\ref{6}), one obtains the full preselected 
distribution function, Eq. (\ref{3}). This result \cite{project} confirms 
that the procedure is a ``true'' post-selection in the Bayesian sense, using 
the channel information to cut the slightly asymmetric preselected distribution 
into two strongly asymmetric postselected ones, which is the basis of signal 
amplification \cite{bayes,diosiprivate}.

Normalizing Eq. (\ref{6}) to the postselected fraction of the incoming
particles, we obtain the postselected probability distribution function
\be{}\label{7}
p_{post}(z)=\frac{q(z)}
  {|\alpha|^2|\gamma|^2+|\beta|^2|\delta|^2
      +2\Re (\alpha\beta^*\gamma^*\delta)e^{-\frac{d^2}{2w^2}}},
\ee
where the denominator is the fraction of all particles postselected into 
spin state $\begin{pmatrix}\gamma\\\delta \end{pmatrix}$, irrespective of 
their displacement $z$.

From Eqs. (\ref{6}) and (\ref{7}) one immediately reads the postselected 
mean displacement
\be{}\label{8}
\begin{split}
\langle z \rangle_{post}
    ~=~&\frac{|\alpha|^2|\gamma|^2-|\beta|^2|\delta|^2}
    {|\alpha|^2|\gamma|^2+|\beta|^2|\delta|^2
      +2\Re (\alpha\beta^*\gamma^*\delta)e^{-\frac{d^2}{2w^2}}}~d
  \\  ~\approx~&\frac{\left(|\alpha|^2|\gamma|^2
                -|\beta|^2|\delta|^2\right)~(d/w)}
    {|\alpha\gamma^*+\beta\delta^*|^2 
     -\Re (\alpha\beta^*\gamma^*\delta)~(d/w)^2}~w.
\end{split}\ee
In the last step we have expanded the exponential to first order in 
$(d/w)^2$, justified by the weakness of the first step of Stern-Gerlach 
separation.

Equation (\ref{8}) is the starting point for the forthcoming discussion. 
First of all, we note that if the measurement is weak enough to satisfy
\be{}\label{9}
\frac{d^2}{w^2}\ll\frac {|\alpha\gamma^*+\beta\delta^*|^2}
                    {|\Re (\alpha\beta^*\gamma^*\delta)|},
\ee
then the post-selected mean value reduces to
\be{}\label{10}
\langle z \rangle_{post}~\approx~\Re\langle z \rangle_{weak}, 
\ee
where 
\be{}\label{11}
\langle z \rangle_{weak}
  =\frac{\alpha\gamma^*-\beta\delta^*}{\alpha\gamma^*+\beta\delta^*}~d
  =\frac{\langle f| \hat\sigma_z d |i\rangle} {\langle f | i \rangle}
\ee
is just the Aharonov-Albert-Vaidman weak value (Eq. (\ref{1})) of the 
observable $\hat A=\hat\sigma_z d$.

Instead of focusing on the shift in coordinate, recent optical 
experiments \cite{kwiat,dixon} measure the shift in transverse momentum 
$\hat p_z$, brought about by the phase difference introduced on 
post-selection. The corresponding mean value can be directly evaluated, 
to give
\be{}\label{12}
\begin{split}
\langle p_z \rangle_{post}
    ~=~&\frac{\Im (\alpha\beta^*\gamma^*\delta)e^{-\frac{d^2}{2w^2}}}
    {|\alpha|^2|\gamma|^2+|\beta|^2|\delta|^2
      +2\Re (\alpha\beta^*\gamma^*\delta)e^{-\frac{d^2}{2w^2}}}
       ~\frac{\hbar d}{w^2}
  \\  ~\approx~&\frac{2~\Im (\alpha\beta^*\gamma^*\delta)~(d/w) }
    {|\alpha\gamma^*+\beta\delta^*|^2 
     -\Re (\alpha\beta^*\gamma^*\delta)~(d/w)^2}
       ~w_p,
\end{split}\ee
where $w_p=(\hbar/2)/w$ is the width of the gaussian momentum distribution 
in the incoming beam. In the limiting case of Eq. (\ref{9}), Eq. (\ref{12}) 
reduces to
\be{}\label{13}
\langle p_z \rangle_{post}~
         \approx~w_p~\Im\langle z \rangle_{weak}~\frac{d}{w}. 
\ee

 Satisfying the condition formulated in Eq. (\ref{9}) is far from being 
trivial though: on approaching post-selection into a spin state 
orthogonal to the preselected one, viz.
\be{}
\label{14}
\gamma\approx i\beta^*;~~\delta\approx -i\alpha^*,
\ee
which may look advantageous in enhancing the signal amplification factor, 
the numerator of the r.h.s. in the inequality (\ref{9}) approaches 
zero. Then the $(d/w)^2$ terms in the denominators of Eqs. (\ref{8}) 
and (\ref{12}) are no more negligible, and it is getting necessary 
to use the full derivative-Lorentzian shape for data fitting. 

The rest of this note addresses that task. Deviations from the 
limiting case of Eq. (\ref{14}) are explored in the following 
parametrization: we introduce the amplitude detuning $\epsilon$ 
through
\be{}
\label{15}
|\gamma|^2=|\beta|^2+\epsilon;~~|\delta|^2=|\alpha|^2-\epsilon
\ee
as well as the phase detuning \cite{thetadelta}
\be{}
\label{16}
\Delta=(\pi-\varphi_{\alpha}+\varphi_{\beta}+
   \varphi_{\gamma}-\varphi_{\delta})/2
\ee
where e.g. $\varphi_{\alpha}$ is the phase angle of the complex
amplitude $\alpha$. Both $\epsilon$ and $\Delta$ are under control 
of the experimenter, fixed for a given run; $d/w$ is the output. 
Expanding the full expressions, Eqs. (\ref{8}) and (\ref{12}) 
to leading powers in $\epsilon$ and $\Delta$, we obtain the 
final results
\be{}
\label{17}
\begin{split}
\langle z\rangle_{post}&=w~\frac
  {2\epsilon}{2|\alpha|^2|\beta|^2+\epsilon(|\alpha|^2-|\beta|^2)}
  \\& \cdot\frac{d/w}{4\Delta^2
       +\left(|\alpha|^{-4}+|\beta|^{-4}\right)(\epsilon^2/8)
        +(d/w)^2}~;
\end{split}\ee
\be{}
\label{18}
\begin{split}
\langle p_z\rangle_{post}&=
 \\w_p&\cdot\frac{2\Delta~(d/w)}{4\Delta^2
       +\left(|\alpha|^{-4}+|\beta|^{-4}\right)(\epsilon^2/8)
        +(d/w)^2}~.
\end{split}\ee

From the above results we see that the fully linear amplification of 
the weak Stern-Gerlach displacement $d$ into the AAV-postselected
mean value, be it real or imaginary, is violated as soon as the 
relative displacement $\eta=d/w$ reaches the order of magnitude of 
\be{}
\label{19}
\eta_0=\left[4\Delta^2
       +\left(|\alpha|^{-4}+|\beta|^{-4}\right)(\epsilon^2/8)\right]^{1/2}.
\ee
Actually, viewed as a function of $\eta$, the value $\eta=\eta_0$ corresponds 
to the maximum of post-selected deflection. Confirming what was expected from 
the phase contrast context, that maximum deflection is limited by the
transverse width of the incoming beam, either in coordinate or in momentum.

For a given value of the weak displacement $d$ to be measured, 
the AAV limit is obtained for  detunings $\Delta$ and/or $\epsilon$ large 
enough to assure $\eta_0~\gg ~d/w$; otherwise the full derivative-Lorentzian
form of Eqs. (\ref{17}) and (\ref{18}) should be used for data fitting,
allowing one to reach higher levels of signal amplification.
It may prove advantageous to build in a controlled value of 
$\epsilon\not=0$, to eliminate errors from unwanted post-selection 
amplitude asymmetry.

The results presented above refer to gaussian beams. As noticed by Hosten 
and Kwiat \cite{kwiat}, switching to momentum space is sensitive to details 
of the beam profile; however, the same can be true about the whole 
calculation \cite{vaidman}. In the one-dimensional case, adding some 
non-gaussian correction terms from a cumulant expansion would be 
straightforward; for real beams of two-dimensional cross-section the
procedure would be more complex. As a general trend, if measured mean 
values are calibrated to the zero deflection case, the remaining dominant 
corrections would appear in the overlap integrals replacing the last term 
in Eq. (\ref{6}), the consequence being a shift of the turning point $\eta_0$ 
from the value given by Eq. (\ref{19}).

We conclude the analysis by noting that the whole AAV procedure can be
regarded as a late descendent of phase-contrast microscopy \cite{zernike}.
In both cases two overlapping  components of slightly different 
space-dependent amplitudes are forced to exhibit a visible interference
pattern, by introducing a constant phase difference between them \cite{tunnel},
thereby turning potentially invisible coherence into large-scale visible 
interference. 

The quantum case is complicated by the fact that the two  overlapping beams 
are entangled to orthogonal spin components which makes their coherence 
strictly latent for straightforward interference counting \cite{latent}.
Post-selecting a spin component corresponding to a different space direction
on the finite-dimension basis of spin states is an ingenious tool to introduce
a phase difference and thereby creating visible phase contrast \cite{geomphase}.
The light polarization case is analogous, although more classical in context: 
even light waves of orthogonal linear polarizations do interfere to produce 
circularly polarized waves; however, that interference remains invisible until 
analyzed (postselected) according to circular polarization. Rotating the basis 
of some other internal degree of freedom can be used to the same end; more 
sophisticated schemes are possible too, like that of two-photon entanglement 
\cite{pryde}. The tool of spin basis rotation followed by selecting one of the 
components is by no means limited to the context of weak measurements; it is 
used for the preparation of so-called Schr\"odinger cat states in mesoscopic 
systems, which is an important step towards quantum information processing 
\cite{cats,eraser}. A more detailed overview of the fascinating field of 
quantum phase contrast physics will be presented elsewhere \cite{AJP}.

A last issue of principle should be mentioned here. There is a growing culture
of theories of weak measurements in a slightly different sense: accomplished 
but unsharp measurement of a ``pointer variable'' coupled to $\hat A$ 
(in our case, the pointer variable is $z$); in a certain limit, that 
unsharp measurement is carried out continuously in time \cite{jacobs,diosi}. 
In the AAV context, a single act of weak measurement in that sense can 
be followed by quantum post-selection according to sharp measurement of 
a different pointer variable (here: $\xi$), the result being used to 
retain or reject data collected during the first step \cite{diosi}. 
A promising implementation is the case of a double-quantum-dot qubit, 
on which the binary position of the electron is measured by means of a 
quantum point contact detector, first weakly, then in a strongly 
post-selecting way \cite{jordanwilliams}. The analysis 
furnishes formulas similar to ours; however, the relationship of the two 
problems to each other may need further clarification.

To summarize, the present note offers a clear and in the same time practical 
framework to discuss pre- and postselected weak measurements. As a first step 
in utilizing the advantages of the proposed approach, we have
started exploring the parameter range next to orthogonal post-selection, 
far beyond the applicability of the Aharonov-Albert-Vaidman weak value, 
which is both accessible and important for signal amplification type 
experiments. 

This work was partially supported by the Hungarian Scientific Research 
Fund OTKA under Grant No. T75129. The author is indebted to Lajos Di\'osi
for enlightening discussions, and to J. Zsolt Bern\'ad, Lev Vaidman, 
Andrew Jordan, Nathan Williams, and Alonso Botero for helpful comments 
on various drafts of the manuscript.


\begin{thebibliography}{99}

\bibitem{aav} Y. Aharonov, D.Z. Albert, and L. Vaidman, Phys. Rev. Lett. 
{\bf 60,} 1351 (1988). For a recent detailed discussion, see the last
three chapters in Y. Aharonov and D. Rohrlich, \textit{Quantum Paradoxes} 
(Wiley-VCH, Weinheim 2005; a most recent brief review is that of 
S. Popescu, Physics {\bf 2,} 32 (2009).
\bibitem{berry} M.V. Berry, in \textit{Quantum Coherence and Reality; 
in celebration of the 60th Birthday of Yakir Aharonov,} J S Anandan 
and J L Safko, eds. (World Scientific, Singapore 1994), p. 55.
\bibitem{ritchie} N.W.M. Ritchie, J.G. Story, and R.G. Hulet, Phys. Rev. 
Lett. {\bf 66,} 1107 (1991)
\bibitem{pryde} G.J. Pryde, J.L. O'Brien, A.G. White, T.C. 
Ralph, and H.M. Wiseman, {\sl ibid.} {\bf 94,} 220405 (2005).
\bibitem{von} One may find it somewhat strange that it is the two 
{\sl unitary} steps which are usually called ``Von Neumann 
{\sl measurements}''.
\bibitem{detectors} Detectors being big, what we have in mind here is 
a detector with a narrow slit in front, scanning together along $z$, 
or a high-resolution detector array.
\bibitem{kwiat} O. Hosten and P. Kwiat, Science {\bf 319,} 787 (2008).
\bibitem{dixon} P.B. Dixon, D.J. Starling, A.N. Jordan, and J.C. Howell, 
Phys. Rev. Lett. {\bf 102,} 173601 (2009).
\bibitem{ampl} It is worth mentioning that signal amplification through
post-selection is a genuine experimental advantage; to achieve the same
enhancement of accuracy by more extensive data acquisition might require
an unrealistic level of temporal stability of the measuring apparatus.
\bibitem{project} More than a mathematical accident, there is a general 
algebraic reason behind: step Eq. (\ref{5}) contains two complementary 
projections in the spin subspace, adding up to unity even if squared. 
\bibitem{bayes} That has been already noticed by Y. Aharonov and A. Botero, 
Phys. Rev. A {\bf 72,} 052111 (2005), who carry on their analysis in terms
of the interesting concept of ``quantum averages of weak values''.
\bibitem{diosiprivate} This also indicates that measurement of $z$, and
post-selection in $\xi$, are interchangeable (L. Di\'osi, private
communication). The nontrivial point about it is that the relative phases 
carried by the spin amplitudes and rendered visible by post-selection, 
being the same for each $z$, are not lost on measuring $z$.
\bibitem{thetadelta} With the factors $i$ in Eq. (\ref{14}), our 
definition of $\Delta$ coincides with that of Hosten and Kwiat 
\cite{kwiat}.
\bibitem{vaidman} L. Vaidman, private communication. 
\bibitem{zernike} F. Zernike, Physica {\bf 9,} 686, 974 (1942).
\bibitem{tunnel}  The most straightforward appearance of that idea in 
quantum mechanics is the tunneling current, induced by the Josephson 
phase difference between two oppositely oriented evanescent waves.
\bibitem{latent} The phrase ``latent coherence'' has been apparently
first used in physics by D.M. Greenberger and A. YaSin, Found. Phys. 
{\bf 19,} 679 (1988).
\bibitem{geomphase} Alternatively, that phase difference can be 
interpreted in terms of a geometrical phase; see S. Tamate, H. Kobayashi,
T. Nakanishi, K. Sugiyama, and M. Kitano, New J. Phys. {\bf 11,} 093025 
(2009).
\bibitem{cats} B. Yurke and D. Stoler, Phys. Rev. Lett. {\bf 57,} 13 (1986); 
C. Monroe, D.M. Meekhof, B.E. King, and D.J. Wineland, Science {\bf 272,} 1131 
(1996); M. Brune {\sl et al.}, Phys. Rev. Lett. {\bf 77,} 4887 (1996).
\bibitem{eraser} Another related emblematic post-selection protocol, 
the ``quantum eraser'', first proposed by M.O. Scully and K. Dr\"uhl, 
Phys. Rev. A {\bf 25,} 2208 (1982), acts just the opposite way. To adapt 
the idea to the present case: destructive interference is first made 
latent by spin labeling, then recovered by projecting spins onto uniform, 
i.e. non-labeling basis states.
\bibitem{AJP} T. Geszti, in preparation.
\bibitem{jacobs} For a recent review, see K. Jacobs and D.A. Steck, 
Contemp. Phys. {\bf 47,} 279 (2006).
\bibitem{diosi} The combination of unsharp quantum measurement and
sharp post-selection is analyzed in detail by L. Di\'osi, in Encyclopedia 
of Mathematical Physics, edited by J.-P. Fran\c coise, G.L. Naber, 
and S.T. Tsou (Elsevier, Oxford 2006), vol. 4, p. 276.
\bibitem{jordanwilliams} N.S. Williams and A.N. Jordan, Phys. Rev.
Lett. {\bf 100,} 026804 (2008); A. Romito, Y. Gefen, and Y.M. Blanter,
{\sl ibid.} {\bf 100,} 056801 (2008).
\end{thebibliography}
\end{document}